\documentclass{arxiv}
\begin{document}

\catchline{}{}{}{}{}

\title{Pion and muon mass difference: a determinning factor in elementary particle mass distribution
}

\author{G. N. Shah$^{*}$ and T. A. Mir$^{**}$}

\address{Nuclear Research Laboratory, Bhabha Atomic Research Centre, \\
Zakura, Srinagar-190 006, Jammu and Kashmir, India\\
$^{*}$drgnshah@rediffmail.com, $^{**}$taarik4@yahoo.co.in}

\maketitle


\section{Abstract}
The most fundamental to the elementary particle is the mass they posses and it would be of importance to explore a possible relationship amongst their masses. Here, an attempt is made to investigate this important aspect irrespective of their nature or scheme of classification. We show that there exists a striking tendency for successive mass differences between elementary particles to be close integral/half integral multiple of the masss diffrence between a neutral pion and a muon. Thus indicating discreteness in the nature of the mass occuring at the elementary particle level. Further, this mass difference of 29.318 MeV is found to be common to the mass spectra of leptons and baryons, implying thereby existance of a basic mechanism linking elementary particles responding to different interactions. 

\section{Keywords}

Elementary particcles; mass relations; mass quantization.


\section{Introduction}	

Explaining the nature and generation of the observed masses of the elementary particles remains an outstanding probem in the modern physics. Experimental observations reveal a wide distribution in the observed masses of the elementary particles with no apparent preferred order. Some studies have related the elementary particle masses to their established quantum numbers whereas others have been mere empirical examimination of their masses in search of some general systematics \cite{Sternheimer1}$^{-}$\cite{Jacobson}. Boson and fermion masses were found to be respectively integral and half integral multiple of $\frac {1}{\alpha}$ electron masses, $\alpha$ being the fine structure constant\cite{Nambu}. On the other hand, in the empirical mass formulla of Frosch\cite{Frosch} the elementary particle masses have been obtained through integral multiplication of thrice the mass of electron.  Masses of the some of the earlier known mesons have also been expressed as integral multiples of 70 MeV\cite{Mac Gregor1} and several mass intervals among elementary particles are reported as integral multiples of the pion mass taken approximately as 140 MeV\cite{Mac Gregor2}. Further, masses of strongly interacting particles are reported to be linear combinations of pion and nucleon mass\cite{Sternheimer1}. Existance of some sequences of particles with mass differences as integral multiples of pion mass and in some others as integral multiples of the muon mass has also been reported\cite{Sternheimer2}. Studies on elementary particles have indicated that baryon and lepton masses are odd integral multiples whereas meson masses are even integral multiples of mass unit of about 35 MeV \cite{Palazzi}. Of late the masses of proton, leptons and quarks have been obtained using electron and muon mass and integral multiples of a mass unit close to muon mass \cite{Mac Gregor3}$^{,}$ \cite{Mac Gregor4}. Clearly electron, muon and pion masses have been used as basic units to show possible discreteness of mass and to explore possible relationship between elementary particle masses. The uncertainy in the determination of the elementary particle mass/mass difference in above studies varied in some cases by a few MeV to about 20 MeV or more\cite{Sidharth}$^{,}$ \cite{Mac Gregor1}$^{,}$ \cite{Mac Gregor2}$^{,}$ \cite{Palazzi}. The conjecture that elementary particle masses are quantized as multiples of the pion mass has lead to intresting physical interpretations and consequances. The constituent quark model was an attempt to reconcile the 35 MeV mass multiplicity with the quark model\cite{Mac Gregor5}. This model was shown to be equivalent to the Nambu's empirical mass formulla\cite{Nambu} and both are related to the existance of Dirac unit of magnetic charge \cite{Akers}. Mass formula involving 70 MeV mass quantum has been derived for the leptons from the magnetic self-interaction of the charge and the anomalous magnetic moment of the relativistic electron\cite{Barut1}$^{,}$ \cite{Barut2}. The 35 MeV mass quantization combined with a stability analysis of the elementary particle mass spectrum has revealed the shell structure of hadrons\cite{Palazzi}.

It is clear from above that a general relationship, linking masses of different elementary particles has consequences in the undrstanding of the fundamentals of physics. In the present study the mass quantum is not chosen a priori instead it follows from our specific mode of analysis. Specifically, elementary particles have been arranged in the  order of their associated physical mass irrespective of their classification and attempt has been made to explore a general relationship the masses of elementary particles may have with each other. We report a significant tendency for mass difference between elementary particles to be close integral/half integral multiple of mass difference between first two massive elementary particles i.e. a neutral pion and a muon.

\section{Data Analysis and Results}

The database for the present study is the latest Particle Data group listings\cite{Eidleman}. Except for the electron all classes of massive particles decaying by weak and/or electromagnetic inteaction i.e. leptons, hadrons and massive gauge bosons have been considered for the analysis. The resonances decaying through strong interaction are excluded from this study due to large uncertainty in their estimated masses\cite{Eidleman}.

\subsection{Leptons,mesons and baryons}

In Table 1 leptons,mesons, baryons and gauge bosons are tabulated in the ascending order of their physical mass irrespective of their structure, type of interaction or their associated quantum numbers. For example the first particle in the table is $\mu^{-}$ a charged lepton followed by $\pi^{0}$ a neutral meson. Similarly, neutron a non-strange baryon is followed by a strangeness '-1' baryon i.e. a lambda particle and strangeness '-3' baryon i.e. omega particle is followed by a tau lepton. In order to look for any specific order in the occurence of these mass states differences between successive masses i.e. $\Delta$$m_{o}$'s were computed and are shown in coloumn 4 of the table 1. On inspection of $\Delta$$m_{o}$'s it was observed that these mass differences could be classified into two categories  Those with values of the order of a few MeV or less were classified as low mass differences and the remaining as the high mass differences. Out of a total of 34 particles leading to 33 mass differences, twenty three were having high mass differences and eleven were having low mass difference. Further, it was observed that all the low mass differences were because of the difference in masses of the different charge states of the same particle and are thus explainable on the basis of electromagnetic interaction\cite{Perkins}. For example 1.3425 MeV is the mass difference between a neutron an da proton and 4.59 MeV is the difference between mass of a charged pion and mass of a neutral pion.

However, the high $\Delta$$m_{o}$'s form the centre of the present investigation as they reflect the mass difference between different types of particles. A closer look at these mass differences reveals that about 12 of these differences are close multiples of 29.318 MeV (Coloum 5), which is the mass difference between $\pi^{0}$ and $\mu^{-}$, the first two particles in the ascending order of physical mass. For example, the observed mass difference between $D^{0}$ a meson and  $\tau^{-}$ a lepton is 87.61 MeV. This value is very close to 87.954 MeV, a value obtained on multiplication of 29.318 MeV by an integer i.e. by 3. Similarly 117.381 MeV, the observed mass difference between two baryons $\Xi^{0}$ and $\Sigma^{-}$ differs from the predicted value of 117.272 MeV by 0.109 MeV. Same is true of the mass difference between pairs of particles ($k^{\pm}$ \& $\pi^{\pm}$), ($\Lambda^{0}$ \& n), ($\Omega^{-}$ \& $\Xi^{-}$), ($D^{*\pm}_{s}$ \& $D^{\pm}_{s}$), ($\Lambda_{c}^{+}$ \& $D^{\pm}_{s}$), ($\Xi_{c}^{+}$ \& $\Lambda_{c}^{+}$), ($\Omega_{c}^{0}$ \& $\Xi_{c}^{/+}$), ($B^{\pm}$ \& $\Omega_{c}^{0}$), ($W^{\pm}$ \& $B_{c}^{\pm}$) and (Z \& $W^{\pm}$) which are very close to the integral multiples of 29.318 MeV and lead to small departures of 2.291, 0.21,  0.676, 2.79, 3.108, 5.492, 1.428, 1.516, 2.95 and 2.894 MeV respectively between the observed and those expected on the basis of integral multiplication of 29.318 MeV. In fact, the uncertainty in the observed masses of $B_{c}^{\pm}$ , $W^{\pm}$ and Z are much larger than the deviations in the mass difference between  observed and predicted values of ($W^{\pm}$ \& $B_{c}^{\pm}$) and (Z \& $W^{\pm}$). The difference between the observed and predicted values in case of particles  ($\eta$ \& $k^{0}$), (p \& $\eta$), ($\Sigma^{+}$ \& $\Lambda^{0}$), ($\tau^{-}$ \& $\Omega^{-}$), ($D^{\pm}_{s}$ \& $D^{\pm}$), ($\Xi_{c}^{/+}$ \& $\Xi_{c}^{0}$), ($B^{*}$ \& $B^{0}$), ($B_{s}^{0}$ \& $B^{*}$), ($\Lambda_{b}^{0}$ \& $B_{s}^{0}$) and ($B_{c}^{\pm}$ \& $\Lambda_{b}^{0}$) are however large and are 8.534, 9.388, 14.264, 12.732, 10.946, 14.346, 13.036, 14.036, 9.462 and 13.732 MeV respectively. It is important to note that seven of these ten large differences are very close half integral multiples of 29.318 MeV and lead to difference between observed and expected values as 0.39, 1.927, 3.713, 0.313, 1.623, 0.623 and 0.927 MeV respectively (column 8). For example the difference between mass of $B^{*}$ and mass of $B^{0}$ differs from the closest integral multiple of 29.318 MeV by 13.036 MeV whereas the difference between the observed and predicted value obtained by the half integral multiplication of mass difference between $\pi^{0}$ and $\mu^{-}$ is only 1.623 MeV. Further, the difference between mass of $\Xi_{c}^{/+}$ and mass of $\Xi_{c}^{0}$ differs from the closest integral multiple of 29.318 MeV by 14.346 MeV whereas the difference between the observed and predicted value obtained by the half integral multiplication of 29.318 MeV is only 0.313 MeV. Clearly most of the differences between observed and calculated values are accountable if the mass difference between successive particles is considered to be half integral multiple of the mass difference between a $\pi^{0}$ and a $\mu^{-}$ i.e ($\Delta$$m_{p2}$)= $\frac {N}{2}$($m_{\pi^{0}}$ - $m_{\mu^{-}}$) where N is an integer. It can be clearly seen from the table 1 that in about 56\% of the cases the difference is in the range of -1 to +2 MeV between the observed and the predicted values and about 74\% cases are accountable with deviation of $\pm$3 MeV. Only in four cases the differences are in the range of 5 to 6 MeV. 
\begin{table}[p]
\tbl{The observed successive mass difference of particls as integral/half integral multiple of 29.318 MeV}
{\begin{tabular}{@{}lllllll@{}} \toprule
Particle & Mass & $\Delta$$m_{o}$  & $\Delta$$m_{p1}$  & Obsd - Expd & $\Delta$$m_{p2}$ & Obsd - Expd \\
& (MeV) & (MeV) & $N$  & (MeV) & $N$ & (MeV)\\ \colrule
$\mu^{-}$\hphantom{00} & 105.65836 & 29.318 & 29.318 & 0 & 29.318 & 0\\
& $\pm$0.00009 & & N=1 & & N=2 \\
$\pi^{0}$\hphantom{00} & 134.9766  & 4.59 &  \\
& $\pm$0.0006\\
$\pi^{\pm}$\hphantom{00} & 139.57018 & 354.107 & 351.816 & 2.291 & 351.816 & 2.291 \\
& $\pm$0.00035 & & N=12 & & N=24\\
$k^{\pm}$\hphantom{00}   & 493.677   & 4.025 &   \\
& $\pm$0.019\\
$k^{0}$\hphantom{00} &   497.648   & 50.102 & 58.636 & -8.534 & 43.977 & 6.125 \\
& $\pm$0.022 & & N=2 & & N=3\\ 
$\eta$\hphantom{00} &    547.75    & 390.522 & 381.134 & 9.388 & 395.793 & -5.271\\
& $\pm$0.12 & & N=13 & & N=27\\ 
$p$\hphantom{00} &       938.27203 & 1.34253 &  \\
& $\pm$0.00008\\ 
$n$\hphantom{00} &  939.56536 & 176.118 & 175.908 & 0.21 & 175.908 & 0.21\\ 
& $\pm$0.00008 & &  N=6 & & N=12\\ 
$\Lambda^{0}$\hphantom{00} & 1115.683 & 73.69 & 87.954 & -14.264 & 73.295 & 0.39 \\
& $\pm$0.006 & & N=3 & & N=5\\ 
$\Sigma^{+}$\hphantom{00} & 1189.37 & 3.1 &  \\
& $\pm$0.07\\ 
$\Sigma^{0}$\hphantom{00} & 1192.642 & 4.97 &  \\
& $\pm$0.024\\ 
$\Sigma^{-}$\hphantom{00} & 1197.449 & 117.3811 & 117.272 & 0.109 & 117.272 & 0.109 \\
& $\pm$0.030 & & N=4 & & N=8\\ 
$\Xi^{0}$\hphantom{00} & 1314.83 & 6.42 &  \\
& $\pm$0.20\\ 
$\Xi^{-}$\hphantom{00} & 1321.31 & 351.14 & 351.816 & -0.676 & 351.816 & -0.676\\
& $\pm$0.13 & & N=12 & & N=24\\ 
$\Omega^{-}$\hphantom{00} & 1672.45 & 104.54 & 117.272 & -12.732 & 102.613 & 1.927 \\
& $\pm$0.29 & & N=4 & & N=7\\ 
$\tau^{-}$\hphantom{00} & 1776.99 & 87.61 & 87.954 & -0.344 & 87.954 & -0.344\\
& +0.29 & & N=3 & & N=6 \\
& -0.26\\  
$D^{0}$\hphantom{00} & 1864.6 & 4.7 &  \\
& $\pm$0.5\\ 
$D^{\pm}$\hphantom{00} & 1869.4 & 98.9 & 87.954 & 10.946 & 102.613 & -3.713\\
& $\pm$0.5 & & N=3 & & N=7\\
$D^{\pm}_{s}$\hphantom{00} & 1968.3 & 143.8 & 146.59 & -2.79 & 146.59 & -2.79\\
& $\pm$0.5 & & N=5 & & N=10\\
$D^{*\pm}_{s}$\hphantom{00} & 2112.1 & 172.8  & 175.908  & -3.108 & 175.908 & -3.108 \\
& $\pm$0.7 & & N=6 & & N=12 \\
$\Lambda_{c}^{+}$\hphantom{00} & 2284.9 & 181.4 & 175.908 & 5.492 & 175.908 & 5.492 \\ 
& $\pm$0.6 & & N=6 & & N=12\\
$\Xi_{c}^{+}$\hphantom{00} & 2466.3 & 5.5 & \\
& $\pm$1.4\\
$\Xi_{c}^{0}$\hphantom{00} & 2471.8 & 102.3 & 87.954 & 14.346 & 102.613 & -0.313 \\
& $\pm$1.4 & & N=3 & & N=7\\
$\Xi_{c}^{/+}$\hphantom{00} & 2574.1 & 4.7 &  \\
& $\pm$3.3\\
$\Xi_{c}^{/0}$\hphantom{00} & 2578.8 & 118.7 & 117.272 & 1.428 & 117.272 & 1.428 \\
& $\pm$3.2 & & N=4 & & N=8\\ \botrule

\end{tabular} \label{ta1}}
\end{table}
\addtocounter{table}{-1}

\begin{table}[h]
\tbl{(Contd.) The observed successive mass difference of particls as integral/half integral multiple of 29.318 MeV}
{\begin{tabular}{@{}lllllll@{}} \toprule
Particle & Mass & $\Delta$$m_{o}$   & $\Delta$$m_{p1}$  & Obsd - Expd & $\Delta$$m_{p2}$  & Obsd - Expd \\
& (MeV) & (MeV) & $N$ & (MeV) & $N$ & (MeV)\\ \colrule

$\Omega_{c}^{0}$\hphantom{00} & 2697.5 & 2581.5 & 2579.984 & 1.516 & 2579.784 & 1.516\\ 
& $\pm$2.6.& & N=88 & & N=176\\ 

$B^{\pm}$\hphantom{00} & 5279 & 0.4 & \\
& $\pm$0.5.\\
$B^{0}$\hphantom{00} & 5279.4 & 45.6 & 58.636 & -13.036 &  43.977 & 1.623 \\
& $\pm$0.5 & & N=2 & & N=3\\
$B^{*}$\hphantom{00} & 5325 & 44.6 &  58.636 & -14.036 & 43.977 & 0.623 \\
& $\pm$0.6 & & N=2 & & N=3\\
$B_{s}^{0}$\hphantom{00} & 5369.6 & 254.4 &  263.862 & -9.462 &  249.203 & 5.197 \\
& $\pm$2.4 & & N=9 & & N=17\\ 
$\Lambda_{b}^{0}$\hphantom{00} & 5624 & 776 & 762.268 & 13.732 &  776.972 & -0.927 \\
& $\pm$9 & & N=26 & & N=53\\
$B_{c}^{\pm}$\hphantom{00} & 6400 & 74025 &  74027.95 & -2.95 & 74027.95 & -2.95 \\
& $\pm$400 & & N=2525 & & N=5050\\
$W^{\pm}$\hphantom{00} & 80425 & 10762.6 & 10759.706  & 2.894 & 10759.706 & 2.894 \\
& $\pm$380 & & N=367 & & N=734\\
$Z$\hphantom{00} & 91187.6   \\ 
& $\pm$21\\ \botrule
\end{tabular} \label{ta1}}
\end{table}
The important feature of the study is a high degree of agreement between observed and predicted mass differences between the members of the SU(3) baryon octet. As can be seen from the Table 1, the differences of observed and predicted mass differences of ($\Lambda^{0}$ \& n), ($\Sigma^{+}$ \& $\Lambda^{0}$) and ($\Xi^{0}$ \& $\Sigma^{-}$) are all found to be ≤$\leq$ 0.39 MeV on the basis of predicted mass differences as integral/half integral multiple of 29.318 MeV. Further the observed mass difference between the last member of baryon octet $\Sigma^{-}$ and the only non-resonance member of the baryon decuplet $\Omega^{-}$ differs from the integral/half integral multiple of 29.318 MeV by only 0.676 MeV. 
\newpage
\subsection{Leptons}

Of the three leptons i.e. $e^{-}$, $\mu^{-}$, $\tau^{-}$ we exclude $e^{-}$ from the study as its mass is several orders of magnitude lower compared to all other particles. The observed mass interval between the remaining unstable leptons i.e. $\mu^{-}$ and $\tau^{-}$ is 1671.331 MeV. The closest value obtained on integral multiplication (N=57) of 29.318 MeV is 1671.126. Clearly this value is very close to the observed value and differs from it by only 0.205 MeV. 

\subsection{Baryons}
Next we consider baryons that decay by weak and/or electromagnetic interaction. Sixteen of these baryons\cite{Eidleman} when listed in the ascending order of mass lead to nine values of high mass difference. The remaining six low mass differences being due to electromagnetic interaction. In   Fig. 1 we show in the form of a histogram the differences between the observed values and those predicted on the basis of $\frac {N}{2}$($m_{\pi^{0}}$ - $m_{\mu^{-}}$) for these baryons. It is clearly seen that the differences are not uniformly spread; instead a clear central tendency exists with 77\% of the cases explainable within a departure of $\pm1$ MeV. This indicates that mass difference between successive baryons occur as half integral multiples of 29.318 MeV. 
  
In case the mass differences were not integral/half integral multiple of 29.318 MeV, distributions as shown in Fig. 1 would hardly be revealed. 
\begin{figure}[th!]
\centerline{\psfig{file=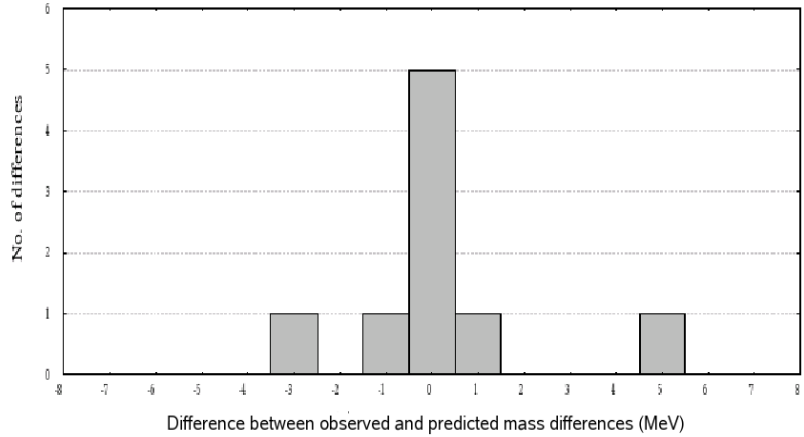,width=5in,height=4in}}
\vspace*{8pt}
\caption{Successive baryon mass differences as half integral multiple of 29.318 MeV.\protect\label{fig:elm.eps}}
\end{figure}
\newpage
Next we studied each baryon with respect to all the remaining baryons by arranging them in the ascending order of mass. The analysis is detailed in Tables 2 to 10.  For example column 2 in Table 2 gives the observed mass difference of proton with respect to each of the remaining baryons. The third column gives the closest value to the observed mass difference obtained on integral multiplication of 29.318 MeV (column 4). A closer look at these predicted mass differences reveals that except for two values all the observed mass differences are close integral multiples of 29.318 MeV. The fifth column is the difference between the observed mass and the closest value obtained on integral multiplication of 29.318 MeV. For example, the observed mass difference between a proton and a lambda particle ($\Lambda^{0}$) is 177.410 MeV. This value is very close to 175.908 MeV, a value obtained on multiplication of 29.318 MeV by an integer 6. Similarly 1759.2279 MeV, the observed mass difference between $\Omega_{c}^{0}$ and proton differs from the predicted value of 1759.08 MeV by 0.1479 MeV. As seen from the tables 2-10 the difference between the observed and the predicted mass difference in case of differences of $\Lambda_{b}^{0}$ with other baryons are somewhat large but are less than the uncertainty ($\pm$9 MeV) in the measured mass of $\Lambda_{b}^{0}$. In some cases the difference of the observed and the predicted mass differences are smaller than the uncertainty in the measured mass of the particles involved. For example, in the mass difference tables for $\Xi_{c}^{/+}$, $\Xi_{c}^{0}$ and $\Omega_{c}^{0}$ the deviation of the predicted mass differences from the observed mass differences are closer to or smaller than the uncertainty (Table 1 Column 2) in the measured mass of the particles involved. 

Interestingly, the only two large differences i.e. for (p \& $\Sigma^{+}$) and ($\Sigma^{+}$ \& $\Lambda^{0}$) between observed and calculated mass differences are accountable if the mass difference ($\Delta$m) between successive particles is considered to be half integral multiple of the mass difference between a $\pi^{0}$ and a $\mu^{-}$ i.e ($\Delta$m)= $\frac {N}{2}$($m_{\pi^{0}}$ - $m_{\mu^{-}}$) where N is an integer. For example the difference between mass of $\Sigma^{+}$ and mass of $\Lambda^{0}$ ( Table 3) differs from the closest integral multiple of 29.318 MeV by 14.267 MeV whereas the difference between the observed and predicted value that would be obtained by the half integral multiplication of the mass difference between $\pi^{0}$ and a $\mu^{-}$ is only 0.392 MeV. Similarly the difference of 12.765 MeV (Table 2) between observed and predicted mass difference of p and $\Sigma^{+}$ differs from the closest half integral multiple of 29.318 MeV by only 1.894 MeV. Thus the mass difference between any two baryons is fully accounted in terms of the mass difference between $\pi^{0}$ a meson and $\mu^{-}$ a lepton . In fact 82\% of the observed 45  (Table 2-10) differences are thus accounted within a deviation of $\pm$4 MeV.  
\newpage
\begin{table}[ph]
\tbl{ Mass difference of baryons with respect to proton as integral multiple of 29.318 MeV}
{\begin{tabular}{@{}lllll@{}} \toprule
P & Mass difference  & $\Delta$m & Integer & Obsd-Expd \\
& (MeV) & $N$($m_{\pi^{0}}$ - $m_{\mu^{-}}$) & N & (MeV)\\ \colrule
$\Lambda^{0}$ \hphantom{00} & 117.410 & 175.908 & 6 & 1.502 \\
$\Sigma^{+}$ \hphantom{00} & 251.097 & 263.862 & 9 & -12.765 \\
$\Xi^{-}$ \hphantom{00} & 383.037 & 381.134 & 13 & 1.903 \\
$\Omega^{-}$ \hphantom{00} & 734.177 & 732.95 & 25 & 1.227 \\ 
$\Lambda_{c}^{+}$ \hphantom{00} & 1346.6279 & 1348.628 & 46 & -2.00\\
$\Xi_{c}^{+}$ \hphantom{00} & 1528.027 & 1524.536 & 52 & 3.491 \\
$\Xi_{c}^{/+}$ \hphantom{00} & 1635.827 & 1641.808 & 56 & -5.981 \\
$\Omega_{c}^{0}$ \hphantom{00} & 1759.2279 & 1759.08 & 60 & 0.1479 \\
$\Lambda_{b}^{0}$ \hphantom{00} & 4685.7279 & 4690.88 & 160 & -5.1521  \\ \botrule
\end{tabular} \label{ta1}}
\end{table}
\begin{table}[ph]
\tbl{ Mass difference of baryons with respect to $\Lambda^{0}$ as integral multiple of 29.318 MeV}
{\begin{tabular}{@{}lllll@{}} \toprule
$\Lambda^{0}$ & Mass difference  & $\Delta$m & Integer & Obsd-Expd \\
& (MeV) & $N$($m_{\pi^{0}}$ - $m_{\mu^{-}}$) & N & (MeV)\\ \colrule
$\Sigma^{+}$ \hphantom{00} & 73.687 & 87.954 & 3 & -14.267 \\
$\Xi^{-}$ \hphantom{00} & 205.627 & 205.226 & 7 & 0.401 \\
$\Omega^{-}$ \hphantom{00} & 556.767 & 557.042 & 19 & -0.275 \\ 
$\Lambda_{c}^{+}$ \hphantom{00} & 1169.217 & 1172.72 & 40 & -3.503 \\
$\Xi_{c}^{+}$ \hphantom{00} & 1350.617 & 1348.628 & 46 & 1.989 \\
$\Xi_{c}^{/0}$ \hphantom{00} & 1463.117 & 1465.9 & 50 & -2.783 \\
$\Omega_{c}^{0}$ \hphantom{00} & 1581.817 & 1583.172 & 54 & -1.355 \\
$\Lambda_{b}^{0}$ \hphantom{00} & 4508.317 & 4514.972 & 154 & -6.665  \\ \botrule
\end{tabular} \label{ta1}}
\end{table}
\newpage
\begin{table}[ph]
\tbl{ Mass difference of baryons with respect to $\Sigma^{-}$ as integral multiple of 29.318 MeV}
{\begin{tabular}{@{}lllll@{}} \toprule
$\Sigma^{-}$ & Mass difference  & $\Delta$m & Integer & Obsd-Expd \\
&(MeV) & $N$($m_{\pi^{0}}$ - $m_{\mu^{-}}$) & N & (MeV)\\ \colrule
$\Xi^{0}$ \hphantom{00} & 117.381 & 117.172 & 4 & 0.109 \\
$\Omega^{-}$ \hphantom{00} & 475.001 & 469.088 & 16 & -5.912 \\ 
$\Lambda_{c}^{+}$ \hphantom{00} & 1078.451 & 1084.766 & 37 & 2.685 \\
$\Xi_{c}^{+}$ \hphantom{00} & 1268.851 & 1260.674 & 43 & 8.177 \\
$\Xi_{c}^{/0}$ \hphantom{00} & 1381.351 & 1377.946 & 47 & 3.405 \\
$\Omega_{c}^{0}$ \hphantom{00} & 1500.051 & 1495.218 & 51 & 4.833 \\
$\Lambda_{b}^{0}$ \hphantom{00} & 4426.551 & 4427.018 & 151 & -0.467 \\ \botrule
\end{tabular} \label{ta1}}
\end{table}
\begin{table}[ph]
\tbl{ Mass difference of baryons with respect to $\Xi^{-}$ as integral multiple of 29.318 MeV}
{\begin{tabular}{@{}lllll@{}} \toprule
$\Xi^{-}$ & Mass difference  & $\Delta$m & Integer & Obsd-Expd \\
&(MeV) & $N$($m_{\pi^{0}}$ - $m_{\mu^{-}}$) & N & (MeV)\\ \colrule
$\Omega^{-}$ \hphantom{00} & 351.14 & 351.816 & 12 & -0.676 \\ 
$\Lambda_{c}^{+}$ \hphantom{00} & 963.59 & 967.494 & 33 & -3.904 \\
$\Xi_{c}^{+}$ \hphantom{00} & 1144.99 & 1143.402 & 39 & 1.588 \\
$\Xi_{c}^{/0}$ \hphantom{00} & 1257.49 & 1260.674 & 43 & -3.184 \\
$\Omega_{c}^{0}$ \hphantom{00} & 1376.19 & 1377.946 & 47 & -1.756 \\
$\Lambda_{b}^{0}$ \hphantom{00} & 4302.69 & 4309.746 & 147 & -7.056 \\ \botrule
\end{tabular} \label{ta1}}
\end{table}
\begin{table}[ph]
\tbl{ Mass difference of baryons with respect to $\Omega^{-}$ as integral multiple of 29.318 MeV}
{\begin{tabular}{@{}lllll@{}} \toprule
$\Omega^{-}$ & Mass difference  & $\Delta$m & Integer & Obsd-Expd \\
&(MeV) & $N$($m_{\pi^{0}}$ - $m_{\mu^{-}}$) & N & (MeV)\\ \colrule
$\Lambda_{c}^{+}$ \hphantom{00} & 612.45 & 615.678 & 21 & -3.288 \\
$\Xi_{c}^{+}$ \hphantom{00} & 793.85 & 791.586 & 27 & 2.264 \\
$\Xi_{c}^{/0}$ \hphantom{00} & 906.35 & 908.858 & 31 & -2.508 \\
$\Omega_{c}^{0}$ \hphantom{00} & 1025.05 & 1026.13 & 35 & -1.08 \\
$\Lambda_{b}^{0}$ \hphantom{00} & 3951.55 & 3957.93 & 135 & -6.38 \\ \botrule
\end{tabular} \label{ta1}}
\end{table}
\newpage
\begin{table}[ph]
\tbl{ Mass difference of baryons with respect to $\Lambda_{c}^{+}$ as integral multiple of 29.318 MeV}
{\begin{tabular}{@{}lllll@{}} \toprule
$\Lambda_{c}^{+}$ & Mass difference  & $\Delta$m & Integer & Obsd-Expd \\
&(MeV) & $N$($m_{\pi^{0}}$ - $m_{\mu^{-}}$) & N & (MeV)\\ \colrule
$\Xi_{c}^{+}$ \hphantom{00} & 181.4 & 175.908 & 6 & 5.492 \\
$\Xi_{c}^{/0}$ \hphantom{00} & 293.9 & 193.18 & 10 & 0.72 \\
$\Omega_{c}^{0}$ \hphantom{00} & 412.6 & 410.452 & 14 & 2.148 \\
$\Lambda_{b}^{0}$ \hphantom{00} & 333.1 & 3342.252 & 114 & -3.152 \\ \botrule
\end{tabular} \label{ta1}}
\end{table}
\begin{table}[ph]
\tbl{ Mass difference of baryons with respect to $\Xi_{c}^{+}$ as integral multiple of 29.318 MeV}
{\begin{tabular}{@{}lllll@{}} \toprule
$\Xi_{c}^{+}$ & Mass difference  & $\Delta$m & Integer & Obsd-Expd \\
&(MeV) & $N$($m_{\pi^{0}}$ - $m_{\mu^{-}}$) & N & (MeV)\\ \colrule
$\Xi_{c}^{/0}$ \hphantom{00} & 112.5 & 117.272 & 4 & -4.772 \\
$\Omega_{c}^{0}$ \hphantom{00} & 231.2 & 234.544 & 8 & -3.344 \\
$\Lambda_{b}^{0}$ \hphantom{00} & 3157.7 & 3166.344 & 108 & -8.644 \\ \botrule
\end{tabular} \label{ta1}}
\end{table}
\begin{table}[ph]
\tbl{ Mass difference of baryons with respect to $\Xi_{c}^{/0}$ as integral multiple of 29.318 MeV}
{\begin{tabular}{@{}lllll@{}} \toprule
$\Xi_{c}^{/0}$ & Mass difference  & $\Delta$m & Integer & Obsd-Expd \\
&(MeV) & $N$($m_{\pi^{0}}$ - $m_{\mu^{-}}$) & N & (MeV)\\ \colrule
$\Omega_{c}^{0}$ \hphantom{00} & 118.7 & 117.272 & 4 & 1.482 \\
$\Lambda_{b}^{0}$ \hphantom{00} & 3045.2 & 3049.072 & 104 & -3.872 \\ \botrule
\end{tabular} \label{ta1}}
\end{table}
\begin{table}[ph]
\tbl{ Mass difference of baryons with respect to $\Omega_{c}^{0}$ as integral multiple of 29.318 MeV}
{\begin{tabular}{@{}lllll@{}} \toprule
$\Omega_{c}^{0}$ & Mass difference  & $\Delta$m & Integer & Obsd-Expd \\
&(MeV) & $N$($m_{\pi^{0}}$ - $m_{\mu^{-}}$) & N & (MeV)\\ \colrule
$\Lambda_{b}^{0}$ \hphantom{00} & 2926.5 & 2931.8 & 100 & -5.3 \\ \botrule
\end{tabular} \label{ta1}}
\end{table}
\section{Discussion}
The elementary particle mass intervals for mesons, baryons and lepton families have been expressed as multiples of basic pion mass \cite{Mac Gregor1}. However, there is no general agreement whether this basic mass is 137 MeV i.e. average of neutral and charged pion mass or some other basic value such as 140 MeV i.e. mass of the charged pion. While the mass unit of 137 MeV reproduces some of the observed mass intervals among the mesons, for the baryons the basic unit of 140 MeV is shown to give a better agreement with the data and for the unstable lepton mass interval the basic unit is reported as 139.7 MeV \cite{Mac Gregor1}. However, if the study is extended to the presently known masses of leptons the predicted mass difference of Ref. \refcite{Mac Gregor1} deviates from the observed value by 5.06 MeV. Again the calculated mass of tau lepton 1786.16 MeV obtained by taking integral i.e. 17 leaps from the muon mass in Ref. 17 and 18 differs from the observed mass 1776.99 MeV by 9.17 MeV. On the other hand we show that the observed mass interval of unstable leptons differs from the integral multiple of 29.318 MeV by only 0.205 MeV. This means that the calculated tau lepton mass obtained by taking integral i.e. 57 leaps of 29.318 MeV from the muon mass differs from the observed mass by only 0.205 MeV. Similarly some hadron mass differences have been reported to be integral multiples of the muon mass with a deviation of $\pm$5 MeV \cite{Sternheimer1}$^{,}$ \cite{Sternheimer2}. On the other hand we have shown with a reasonably better accuracy that the mass unit of 29.318 MeV is basic to the mass difference between elementary particles arranged in ascending order of mass. This holds true for mass differences between unstable leptons, successive baryons and of 45 mass differences among all baryons. Such an evidence cannot be mere accidental. The high mass differences between the members of different baryon isospin multiplets compared to small mass intervals within a multiplet are attributed to the higher magnitude of the strong interaction than the electromagnetic interaction. It is therefore intriguing that the high mass differences due to strong interaction like that between members of baryon octet are close integral multiples of a mass difference between strongly interacting pion and electromagnetically/weakly interacting muon. Further, as shown in the present study the mass difference between a neutral pion (hadron) and a muon (lepton), is common to both leptons and baryons. Although leptons and hadrons are treated as unrelated entities, pion and muon mass difference appears to be basic to the mass spectra of these particles and hence bridges the two separate realms of the elementary particles. The general implication of our study is that elementary particle mass states seem quantized and that they occur distributed in various mass states such that the mass differences are integral/half integral multiples of the mass difference between first two lowest massive elementary particles. This goes well with the general rule in physical systems where lowest mass states are considered as the building blocks of more complex systems and hence in some sense the most fundamental. From the study it looks possible to obtain mass of an elementary particle by making integral/half integral jumps in units of 29.318 MeV.  This discreteness of mass occurrence holds fairly well for baryons and leptons. The present study treating the leptons, mesons and baryons on equal basis will have impact in the understanding of the physics of the elementary particles.
\section{Conclusion}
Our study reveals a general tendency for the mass differences between the elementary particles to be integral/half integral multiples of mass difference between a neutral pion and a muon, the first two massive elementary particles occurring in nature. The results are equally applicable to all particles arranged in the ascending order of mass irrespective of their nature or classification, to successive members of the baryon octet, to the unstable leptons and to the mass difference between any two baryons and hence can not be treated as a mere coincidence. That the pion and muon mass difference appears to be basic to the mass spectra of leptons and baryons tends to indicate a sort of common link between particles responding to different interaction forces.

\section*{Acknowledgments}
The authors are highly thankful to Paolo Palazzi, Amjad Hussain Shah Gilani and David Akers for their helpful comments and suggestions on the contents of the paper. 




\begin{thebibliography}{0}
\bibitem{Sternheimer1} R. M. Sternheimer, {\text Phys.  Review Lett. 13}, 358-360 (1964)
\bibitem{Sternheimer2} R. M. Sternheimer, {\text Phys.  Review. 2B}, 138, 446-478 (1965)
\bibitem{Sidharth} B. G. Sidharth, {\text arXiv: physics/0306010}, (2003)
\bibitem{jpap4} Raja Ramanna and B. V. Sreekantan, {\text Mod. Phys. Lett A. 10}, 741-753 (1995)
\bibitem{Gareev} Gareev et al, {\text Acta Physica Polonica B. 29}, 2493-2500 (1998)
\bibitem{Chizov} M. V. Chizov, {\text arXiv: hep-ph/0107025} (2001)
\bibitem{Milikan} R. C. Milikan and D. C. Richman, {\text arXiv: hep-ph/0106106} (2001)
\bibitem{Giani} S. Giani, {\text Report No. CERN-OPEN-2004-004} (2004)
\bibitem{Gilani} A. H. S. Gilani, {\text arXiv: hep-ph/0503196} (2005)  
\bibitem{Jacobson} T. Jacobson, {\text arXiv: hep-ph/0502205} (2005)
\bibitem{Nambu} Y. Nambu, {\text Prog. Th. Phys. 7},  595 (1952)
\bibitem{Frosch} R. Frosch, {\text Nuovo Cimento. 104 A}, 913 (1991)
\bibitem{Mac Gregor1} M. H. Mac Gregor, {\text Lett. Nuovo Cimento. 14}, 211 (1970)
\bibitem{Mac Gregor2} M. H. Mac Gregor, {\text IL Nuovo Cimento. A58}, 159 (1980)
\bibitem{Palazzi} P. Palazzi, {\text http://particlez.org/p3a/abstract/2006-01.htm} (2006)
\bibitem{Mac Gregor3} M. H. Mac Gregor, {\text arXiv: hep-ph/0607233}, (2006)
\bibitem{Mac Gregor4} M. H. Mac Gregor, {\text arXiv: hep-ph/0603201}, (2006) 
\bibitem{Mac Gregor5} M. H. Mac Gregor, {\text IL Nuovo Cimento. 103 A}, 983 (1990) 
\bibitem{Akers} D. Akers, {\text Int. J. Th. Phys. 33}, 1817 (1994)
\bibitem{Barut1} A. O. Barut, {\text Phys. Lett. 73B }, 310 (1978)
\bibitem{Barut2} A. O. Barut, {\text Phys. Rev. Lett. 42}, 1251 (1979)
\bibitem{Eidleman} S. Eidelman et al, {\text Phys. Lett. B 1}, 592 (2004)
\bibitem{Perkins} D. H. Perkins, {\text Introduction to High Energy Physics},pp. 126-129, 4th Edn. (Cambridge University Press, 2000)


\end{thebibliography}
\end{document}